# Limits for the graphene on ferroelectric domain wall p-n-junction rectifier for different regimes of current

Maksym V. Strikha[1,2] and Anna N. Morozovska[3],

[1] *V.Lashkariov Institute of Semiconductor Physics, National Academy of Sciences of Ukraine,*
*pr. Nauky 41, 03028 Kyiv, Ukraine*

[2] *Taras Shevchenko Kyiv National University, Radiophysical Faculty*
*pr. Akademika Hlushkova 4g, 03022 Kyiv, Ukraine*

[3] *Institute of Physics, National Academy of Sciences of Ukraine,*
*pr. Nauky 46, 03028 Kyiv, Ukraine*

**Abstract**

Here we present the theory of the conductivity of pn junction (pnJ) in graphene channel, placed on ferroelectric substrate, caused by ferroelectric domain wall (FDW) for the case of arbitrary current regime: from ballistic to diffusive one. We calculated the ratio of the pnJ conductions for opposite polarities of voltages, applied to source and drain electrodes of the channel, $G_+^{total}/G_-^{total}$, as the function of the graphene channel length $L$, electron mean free path $\lambda$ and ferroelectric permittivity $\varepsilon_{33}^f$. We have demonstrated, that the small values of $G_+^{total}/G_-^{total}$ (0.1 and smaller), which correspond to efficient graphene pnJ based rectifier, can be obtained for the ferroelectrics with high $\varepsilon_{33}^f \gg 100$ and for the ratios of $L/\lambda\sim 1$ or smaller. However, for ferroelectrics with extremely high $\varepsilon_{33}^f$ (relaxor or $PbZr_xTi_{1-x}O_3$ with composition x near the morphotropic phase boundary x=0.52) the ratio $G_+^{total}/G_-^{total}$ can be essentially smaller then unity for the case of a pronounced diffusive regime of current as well. This makes the ferroelectric substrates with high permittivity excellent candidates for the fabrication of new generation of rectifiers based on the graphene pnJ.

Temperature effect on $G_+^{total}/G_-^{total}$ ratio was studied within Landau-Ginzburg-Devonshire approach. We have demonstrated that rectifying properties of the graphene pnJ become better in the vicinity of Curie temperature. However, for the temperatures higher than Curie temperature the rectifying effect vanishes due to the ferroelectric polarization disappearance.

# I. Introduction

The remarkable properties of the p-n-junction (**pnJ**) in graphene were studied in numerous papers since middle 2000[th] [see, e.g. 1, 2, 3, 4, 5]. As it was demonstrated theoretically in [3], the transmission of electron through pnJ potential barrier at an angle normal to the barrier occurs with probability 1. Physically this can be explained by electron wave function structure in graphene: the incoming and the reflected electron states have opposite chirality, which makes the probability of reflection equal to zero [6, 7]. This phenomenon is usually treated as the evidence of relativistic Klein paradox in the experiment on "table desk".

Until recently, pnJ were generally realized by the system of two gates, one of them doping one region of graphene channel with electrons, and the other doping the other region with holes. The pnJ in graphene was experimentally realized by Williams et al [8] and studied theoretically by Zhang and Fogler [2] in a system consisting of a gated graphene channel (**GC**) and a top gate superimposed through the dielectric layer above the GC. It is important, that Zhang and Fogler [2] obtained that an electric field in the pnJ is high in comparison with elementary estimations due to the low screening ability of two-dimensional (2D) Dirac quasiparticles. Hence the resistance of pnJ is quite low. Later on the physical principles of such junctions operation have been explained by Beenakker [9]. Further experimental studies were focused on the quantum Hall effect, the Klein (or Landau-Zener) tunneling and the Veselago lensing in graphene pnJs [10, 11].

For a long time the only way to design pnJs in graphene was the usage of multiple gates or the chemical doping of separate sections of graphene channel, till Hinnefeld et al [12] and Baeumer et al [13] created a pnJ in graphene using the ferroelectric substrates $Pb(Zr,Ti)O_3$ (PZT) and $LiNiO_3$ correspondingly. At that Baeumer et al [13] proposed the pnJ in graphene at a ferroelectric domain wall. The principal idea of the latter work is that if graphene is imposed on a 180°-ferroelectric domain wall (**FDW**), a pnJ can arise without applying any additional gates, doping or screening.

A theoretical model for the electric field and ballistic transport in a single-layer GC at a 180°-FDW have been developed recently [14]. The impact of the FDW on the ballistic conductance of a single-layer graphene channel in the graphene / physical gap / ferroelectric film heterostructure has been studied there in the Wentzel-Kramers-Brillouin (**WKB**) approximation. The self-consistent numerical simulation of the electric field and the space charge dynamics in the heterostructure, and the approximate analytical theory show that the contact between the domain wall and the surface creates a pnJ junction in the GC. We calculated that the carrier concentration induced in graphene by uncompensated ferroelectric dipoles originated from the abrupt spontaneous polarization change near the surface can reach value of about $10^{19}$ m$^{-2}$ for the ferroelectric with substantial spontaneous polarization, which is two orders of magnitude higher than those obtained for the graphene on non-ferroelectric substrates. Therefore, we had predicted that the graphene channel with the p-n junction caused by the 180°-FDW would be characterized by rather a high ballistic conductivity. Moreover, the graphene pnJ at theFDW can be an excellent rectifier with a conductivity ratio between the direct and reverse polarities of applied voltage of about 10.

In this paper we expand the results of [14] to the case of different types of current regime (diffusive, quasi-ballistic and ballistic one) and examine the characteristics of pnJ in graphene at FDW properties as functions of ferroelectric and graphene conducting channel properties, such as ferroelectric permittivity and Curie temperature, and a ratio between the graphene conducting channel length and an electron mean free path in this channel.

## II. Analytical theory

The geometry of the examined system is shown in **Figure 1.**

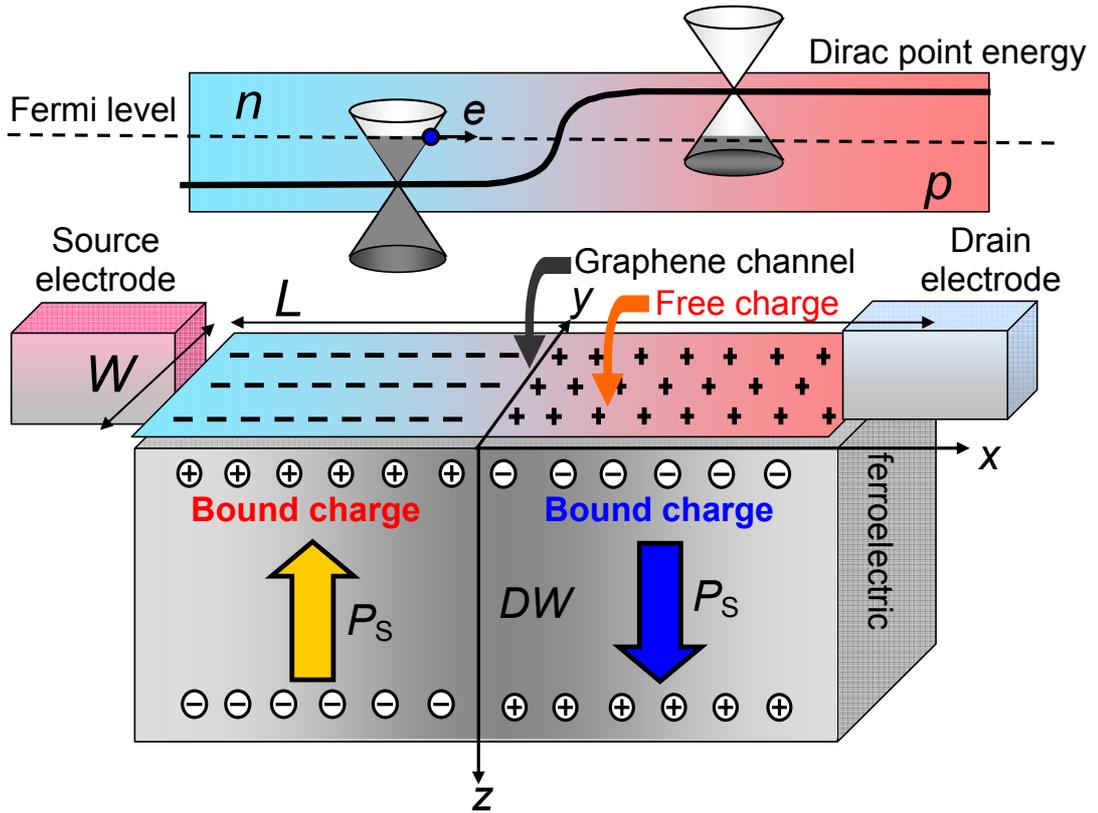

**Figure 1.** 180°-DW structure near the uniaxial ferroelectric surface in the heterostructure graphene-on-ferroelectric (bottom part) and the potential barrier of pnJ for an electron (upper part). Single-layer GC length is $L$, its width is $W$. Ferroelectric substrate with polarizations $\pm P_S$ separated by a 180°-FDW is thick enough.

Discontinuity of the double electric layer consisting of screening and bound charges results in the depolarization electric field that strays into the graphene monolayer. The electrostatic problem of the electric field distribution was solved in ref [14], several modifications have been considered earlier [15, 16, 17].

For the electron, moving from the left part of the channel to the right one, the potential barrier arises with the height equal to the double Fermi energy $E_F$ that corresponds the 2D electrons concentration in graphene $n_{2D}$,:

$$\Delta E = 2E_F = 2\hbar v_F \sqrt{\pi n_{2D}} \tag{1}$$

Here $v_F = 10^6$ m/s is a velocity constant in the expression for graphene linear spectrum $E = \pm\hbar v_F k$, where $k$ is a wave vector [6]. The concentration $n$ is caused by the spontaneous ferroelectric polarization $P_S$ and its permittivity tensor, $n_{2D} \cong 2\gamma(P_S/e)$, where the dielectric anisotropy factor $\gamma \sim 1$ [14]. Hence $k_F = \sqrt{\pi n_{2D}} \sim \sqrt{\pi(P_S/e)}$. Mention that there is no potential barrier at all for the electron moving in opposite direction from the right to the left in **Figure 1**.

The ballistic conduction of the GC can be presented by the most general Landauer formula (see e.g. [18]). In the limit of strongly degenerated carriers, which is obviously valid for graphene, doped by the essential ferroelectric polarization $P_S$, it can be rewritten as:

$$G^{ball} = \frac{2q^2}{h} M(E_F). \tag{2}$$

Here $M(E)$ is a number of conduction modes. For 2D single layer graphene we get [7]:

$$M(E) = \frac{2|E|}{\pi \hbar v_F} W \tag{3}$$

Here $W$ is the graphene channel width in $y$ direction (see **Figure 1**). Allowing for (1), for the form of graphene band spectrum we get:

$$G^{ball} = \frac{4e^2}{h} W k_F \tag{4}$$

Here $k_F$ is the wave vector corresponding to the Fermi energy:

$$k_F = \frac{E_F}{\hbar v_F} = \sqrt{\pi n_{2D}} \tag{5}$$

For the case, when electron is passing through the potential barrier of the pnJ, Eq.(4) is to be multiplied by the transmission probability $w(\vartheta) \sim \cos^2 \vartheta$ [3] and averaged over all the angles $\vartheta$ between $x$ axis along the graphene channel and the direction of electron movement:

$$G^{ball}_{pn} = \frac{4e^2}{h} W \int \frac{k_F w(\vartheta)}{2\pi} d\vartheta \tag{6}$$

As it was demonstrated in [3], Eq.(6) in WKB approximation yields:

$$G_{pn}^{ball} \approx \frac{2e^2}{\pi\hbar} W \sqrt{\frac{k_F}{d}} \qquad (7)$$

Here $d$ is a characteristic length-scale of pnJ, determined by the geometry and the parameters of the problem. As it was shown in [14], for the system, analyzed here and presented in **Figure 1**, we get:

$$G_{pn}^{ball} \cong \frac{e^2}{\pi\hbar} W \sqrt{\frac{\alpha}{\varepsilon_{33}^f} \frac{c}{\pi v_F} n_{2D}} \qquad (8)$$

Here, $\alpha = e^2/4\pi\varepsilon_0(\hbar c)$ is the fine-structure constant, $c$ the light velocity in vacuum, and $\varepsilon_{33}^f$ is the permittivity of ferroelectric along $z$ direction, that is normal to the graphene plane. The comparison of Eqs. (7) and (8) yields

$$d = \frac{1}{k_F} \frac{\pi^2}{2} \frac{\varepsilon_{33}^f v_F}{\alpha c} \approx \frac{2.3\varepsilon_{33}^f}{k_F} \qquad (9)$$

Because of the large value of ferroelectric permittivity (>5000 for ferroelectric relaxor, 500 for Pb(Zr,Ti)O$_3$, 120 for BaTiO$_3$, 50 for LiTaO$_3$ and 36 for LiNbO$_3$ at room temperature) we get $k_F d > 1$ which is a criteria for a smooth pnJ, used as a principal assumption in ref.[3].

Mention that a formalism of Landauer formula (2) needs no concept of holes, used in a common "Drude treatment" [18]. The conduction is determined by a number of conduction modes for electrons in a vicinity of Fermi level only. Therefore a situation is different for opposite polarities, applied to source and drain in a GC, namely the ballistic conductivity for polarity " – " on source and "+" on drain (when electrons are moving in **Figure1** from left to right) is described by Eq.(8). However, for the opposite polarity "+" on source and " – " on drain (when electrons are moving in **Figure 1** from right to left) it is described by Eq.(4), which can be rewritten with allowing for Eq.(5) as:

$$G^{ball} = \frac{2e^2}{\hbar\pi^{3/2}} W \sqrt{n_{2D}} \qquad (10)$$

Therefore the ratio of the two conductions

$$\frac{G_{pn}^{ball}}{G^{ball}} = \frac{\sqrt{\pi}}{2} \sqrt{\frac{\alpha c}{\varepsilon_{33}^f v_F}} \equiv \beta(\varepsilon_{33}^f) \qquad (11)$$

can be of 0.1 order and smaller due to the high value of permittivity $\varepsilon_{33}^f$, that makes a barrier "smoother" due to Eq.(9) and the transmission smaller than for a "sharp" one due to Eq.(7). Note, that a similar effect of the pnJ barrier smoothness or sharpness on its conduction was examined for the first time in [2]. Therefore the pnJ in graphene at FDW can be an excellent rectifier [14].

However, the majority of real graphene devices, especially based on the less perfect CVD graphene, is described by the mean free path of electrons λ (λ ~ 50-250 nm order) and are operating in a diffusive regime [6]. In order to get a conduction value for this case we'll formally divide the graphene channel with the length $L$ between the source and the drain into 3 sections: one, containing pnJ itself, with the length λ, and the two other, on both sides from pnJ, with total length $L - \lambda$ (see Fig.2). We can do this, because in graphene channel the electron mean free path $\lambda$ is generally longer than the intrinsic width of the uncharged domain wall in proper ferroelectric, that is about 1 – 10 nm [19]. For the case $L \gg \lambda$ the current in these two sections flows in diffusive regime.

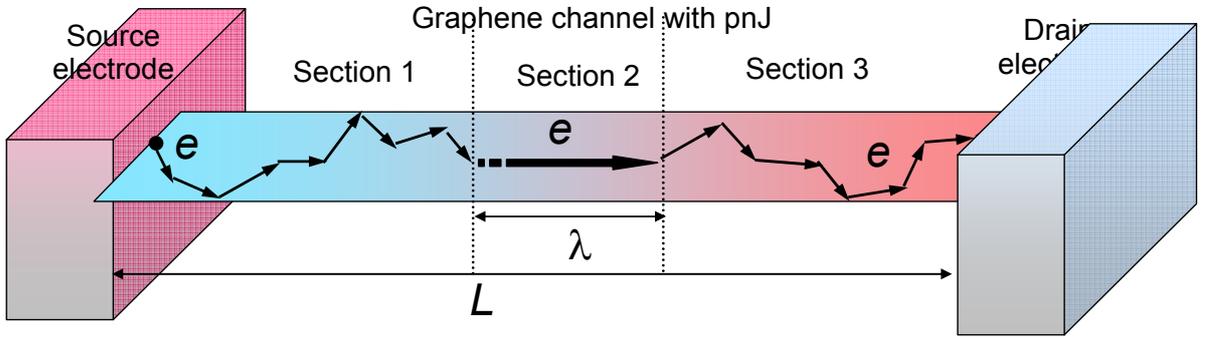

**Figure 2.** Formal division of graphene channel into three sections. The current in Sections 1, 3 is diffusive, while in Section 2 it is ballistic.

However, the current in a narrow central section with pnJ itself can still be treated as a ballistic one, and the conduction of this section can still be described by Eqs. (8), (10) for different polarities of voltage. We'll also examine now the case of small voltage $V$ applied between the source and drain, $eV < k_B T$, where $k_B$ is Boltzmann constant, $T$ is temperature. In the case Fermi levels of both contacts are within the same Fermi window of conductivity with the width in energy scale $\sim 2k_B T$; and so inelastic collisions are not required in order to enable electron reaching the drain after starting at the source. Therefore the total conduction of these two sections can be written as [18]:

$$G^{diff} = \frac{\lambda}{L + \lambda} G^{ball} \qquad (12)$$

where $G^{ball}$ is determined by Eq.(10). Finally the total conduction of the sample is governed by evident expression:

$$\frac{1}{G^{total}} = \frac{1}{G^{diff}} + \frac{1}{G^{ball}} \qquad (13)$$

The change of voltage polarity on contacts does not change the conduction of sections 1 and 3 in any way. The conduction of section 2 is described in one case by Eq.(8), while there is no need for separation of this section in the opposite polarity case, because all the length of GC is characterized by diffusive conduction without any tunnelling. Therefore in general case the ratio of conductions for different polarities with allowing for Eqs. (10) – (13) can be written as:

$$\frac{G_+^{total}}{G_-^{total}} = \frac{\beta(L+\lambda)}{\beta(L+\lambda)+\lambda} \quad (14)$$

Mention, that the parameter $\beta$ depends on the dielectric permittivity, $\beta(\varepsilon_{33}^f) = \sqrt{\pi\alpha c/(4\varepsilon_{33}^f v_F)}$ and the electron mean free path $\lambda$ depends on the concentration $n_{2D}$, and so on ferroelectric polarization, since $n_{2D} \sim \pi(P_S/e)$ [14]. When the scattering of electrons in graphene channel at charged impurities in ferroelectric is dominant (which is a most common case for the real graphene operational device), $\lambda(n) \sim \sqrt{n}$ (see e.g. [6]). For the case of the short ballistic channel, $L << \lambda$, Eq.(14) obviously transforms into Eq.(8). In the opposite limit of the long diffusive channel, $L >> \lambda$, we get

$$\frac{G_+^{total}}{G_-^{total}} = 1 \quad (15)$$

so that the rectifying properties of pnJ vanishes.

Mention, that for the opposite case of comparatively high voltages, $eV >> k_B T$, the proposed consideration is no longer valid. Now there are two Fermi windows of conductivity around two essentially different Fermi levels of source and drain, and therefore an electron needs inelastic collision to transfer in energy scale from one window to other in order to overpass the GC from the source to drain. The consideration based on the framework of ref.[18] leads to the result

$$I = I_o(e^{eV/vk_BT} - 1) \quad . \quad (16)$$

Here $I_o$ and $v$ parameters depend dominantly on the physical nature of elastic collisions.

### III. Results and discussion

The ratio $G_+^{total}/G_-^{total}$ as a function of ratio $L/\lambda$, graphene channel length $L$, and ferroelectric permittivity $\varepsilon_{33}^f$ calculated from Eq.(11) for different ferroelectric substrates is presented in **Figure 3**. Ferroelectric materials parameters are listed in **Table I.**

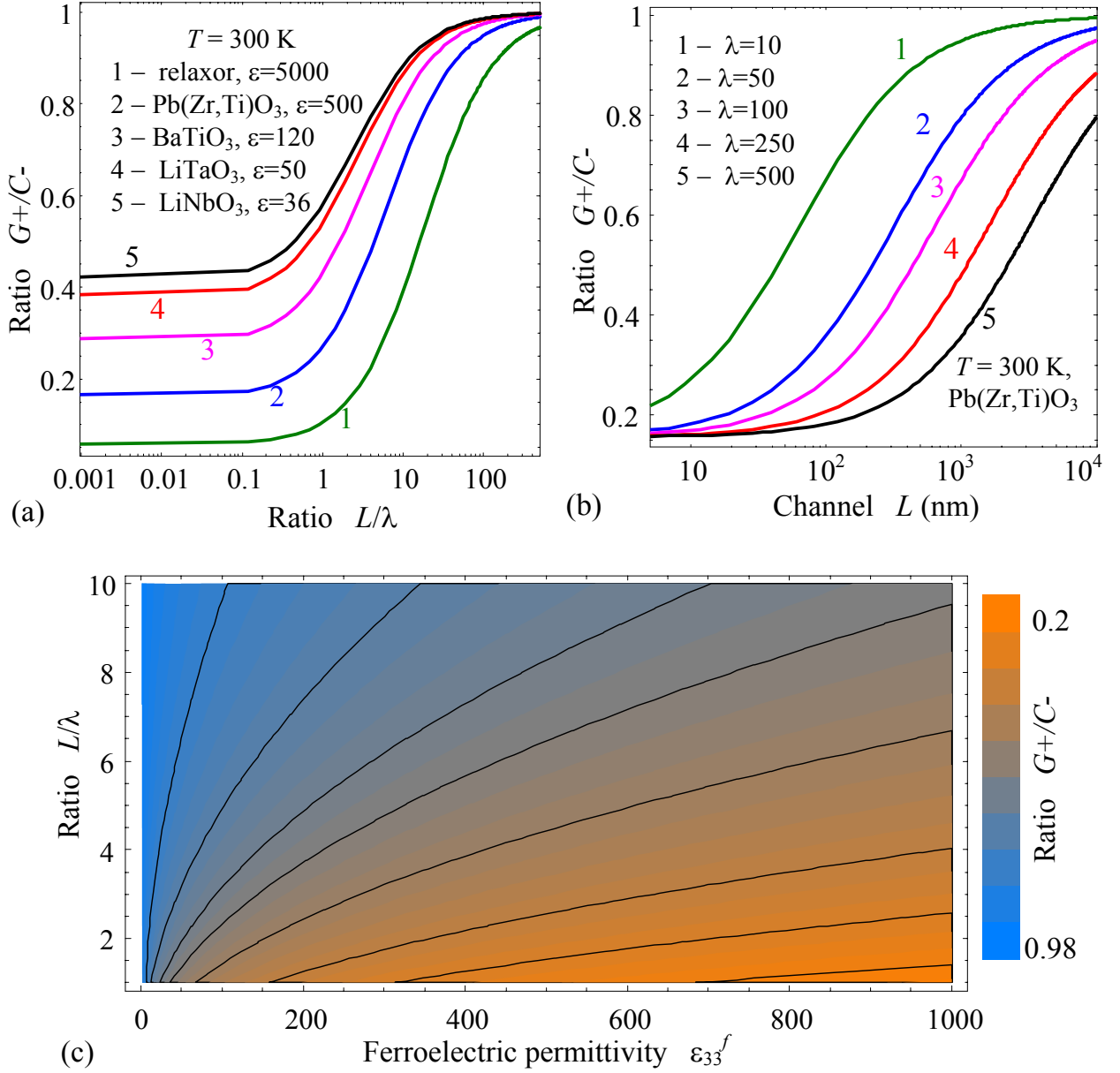

**Figure 3. (a)** The ratio $G_+^{total}/G_-^{total}$ as a function of ratio $L/\lambda$ calculated from Eq.(11) for different ferroelectric substrates, namely for several $\beta(\varepsilon_{33}^f)$, where the dielectric permittivity $\varepsilon_{33}^f \geq 5000$ for ferroelectric relaxor, 500 for Pb(Zr,Ti)O$_3$, 120 for BaTiO$_3$, 50 for LiTaO$_3$ and 36 for LiNbO$_3$ at room temperature 300 K (curves 1-5). **(b)** The ratio $G_+^{total}/G_-^{total}$ as a function of graphene channel length $L$ calculated for several mean paths $\lambda$ = (10, 50, 100, 250, 500) nm (curves 1-5) for ferroelectric Pb(Zr,Ti)O$_3$ substrate and $T$ = 300 K. **(c)** Contour map of the ratio $G_+^{total}/G_-^{total}$ in coordinates $\{L/\lambda, \varepsilon_{33}^f\}$. Ferroelectric materials parameters are listed in **Table I**.

**Table I. Parameters of ferroelectrics**

| Quantity | Ferroelectric PbZr$_x$Ti$_{1-x}$O$_3$ (x ≈ 0.5) (PZT50/50) | Ferroelectric BaTiO$_3$ |
|---|---|---|
| media permittivity | $\varepsilon^b_{33}$=7, $\varepsilon^f_{33}$=500 at room temperature | $\varepsilon^b_{33}$=7, $\varepsilon^f_{33}$=120 at room temperature |
| LG – potential coefficients | $\alpha_T$=2.66×10$^5$ C$^{-2}$·mJ/K,  $T_C$=666 K<br>$b$ = 1.91×10$^8$ J C$^{-4}$·m$^5$<br>$\gamma$ = 8.02×10$^8$ J C$^{-6}$·m$^9$ | $\alpha_T$=6.68×10$^5$ C$^{-2}$·mJ/K,  $T_C$=381 K<br>$b$ = (-8.18×10$^9$+1.876$T$×10$^7$) J C$^{-4}$·m$^5$<br>$\gamma$ = (1.467×10$^{11}$-3.31$T$×10$^8$) J C$^{-6}$·m$^9$. |

As one can see from **Figure 3**, small values of $G^{total}_+ / G^{total}_-$, which correspond to efficient graphene pnJ based rectifier, can be obtained for the the ratios of $L/\lambda$ of unity order or smaller and ferroelectric substrates with high permittivity. The expected result can be very realistic for the channels of sub-micron length in the case of the comparative values of the electron mean free path. The situation corresponds to the current regime transient from diffusive to ballistic one; mention, that the distance between the parallel FDW is generally much longer, in [7] it was of 5μm order, which enables to fabricate a micron-length channel with contacts near one FDW only. However, for ferroelectrics with extremely high permittivity, such as relaxors or PbZr$_x$Ti$_{1-x}$O$_3$ with composition x near the morphotropic phase boundary x=0.52, $G^{total}_+ / G^{total}_-$ ratio can be essentially smaller then 1 for the case of a pronounced diffusive regime of current as well. This allows us to consider ultra-high permittivity ferroelectric substrates as excellent candidates for the fabrication of the novel rectifiers based on graphene pnJ.

The results presented in **Figure 3** were calculated at room temperature. To study the temperature behaviour of the pnJ conductivity, which can be essential for our problem for *T* near the Curie one, we used the Landau-Ginzburg-Devonshire (LGD) type model of ferroelectric substrate, for which the value of the ferroelectric polarization $P_3$ is given by the LGD-equation

$$a(T)P_3 + bP_3^3 + \gamma P_3^5 - g\frac{\partial^2 P_3}{\partial z^2} = E_3 \qquad (17)$$

According to Landau theory [20, 21], the coefficient $a(T) = \alpha_T(T - T_C)$ explicitly depends on temperature *T*, $T_C$ is the Curie temperature. All other coefficients are supposed to be temperature independent. Coefficient *b*>0 for the ferroics with the second order phase transition and *b*<0 for the first order one. Quasi-static electric field is defined via electric potential as $E_3 = -\partial\varphi/\partial x_3$, where the potential φ satisfies a conventional Poisson equation inside a ferroelectric layer. The relevant electrostatic boundary conditions are standard for ferroelectrics systems. The temperature dependences of homogeneous spontaneous polarization and dielectric permittivity are given by expressions:

$$P_S^2(T) = \frac{1}{2\gamma}\left(\sqrt{b^2 - 4a(T)\gamma} - b\right), \quad \varepsilon_{33}^f(T) = \frac{\varepsilon_0}{a + 3bP_S^2(T) + 5\gamma P_S^4(T)} + \varepsilon_b \qquad (18)$$

For the ferroelectric with the second order phase transition to a paraelectric phase ($b > 0$, $\gamma = 0$),

$$P_S(T) = P_S(0)\sqrt{1 - (T/T_C)} \text{ and } \varepsilon_{33}^f(T < T_C) = \frac{\varepsilon_{33}^0}{2|1 - (T/T_C)|} + \varepsilon_b \text{ and } \varepsilon_{33}^f(T > T_C) = \frac{\varepsilon_{33}^0}{|1 - (T/T_C)|} + \varepsilon_b.$$

In a paraelectric phase $P_S = 0$. Note, that temperature dependence of spontaneous polarization affects the concentration of electrons in graphene channel [9], which, in its turn, influences on the Fermi energy and the electron mean free path [6, 7]. From the expressions (18) the coefficient β and mean path λ becomes polarization dependent and so temperature dependent, namely:

$$\beta(T) = \sqrt{\frac{\pi\alpha c}{4v_F \varepsilon_{33}^f(T)}}, \qquad \frac{1}{\lambda(T)} = \frac{1}{\lambda_f}\sqrt{\frac{P_S(0)}{P_S(T)}} + \frac{1}{\lambda_0} \qquad (19)$$

Where the values $\lambda_f$ and $\lambda_0$ are temperature-independent parameters of the theory. At that $\lambda_f$ corresponds the mean free path, determined by scattering at ionized centres in the substrate, in the temperature range, far from Curie temperature, where $\lambda_f(n) \sim \sqrt{n}$ [5], $\lambda_0$ represents a generalized mean free path, determined by rival scattering mechanisms (e.g. scattering an short range imperfections in graphene 2D lattice; it limits the general mean free path when $\lambda_f$ is too long).

In **Figure 4** we present the temperature dependence of the ratio $G_+^{total}/G_-^{total}$ calculated for different values of $\lambda_f$, for the two types of substrates: BaTiO$_3$ and Pb(Zr,Ti)O$_3$.

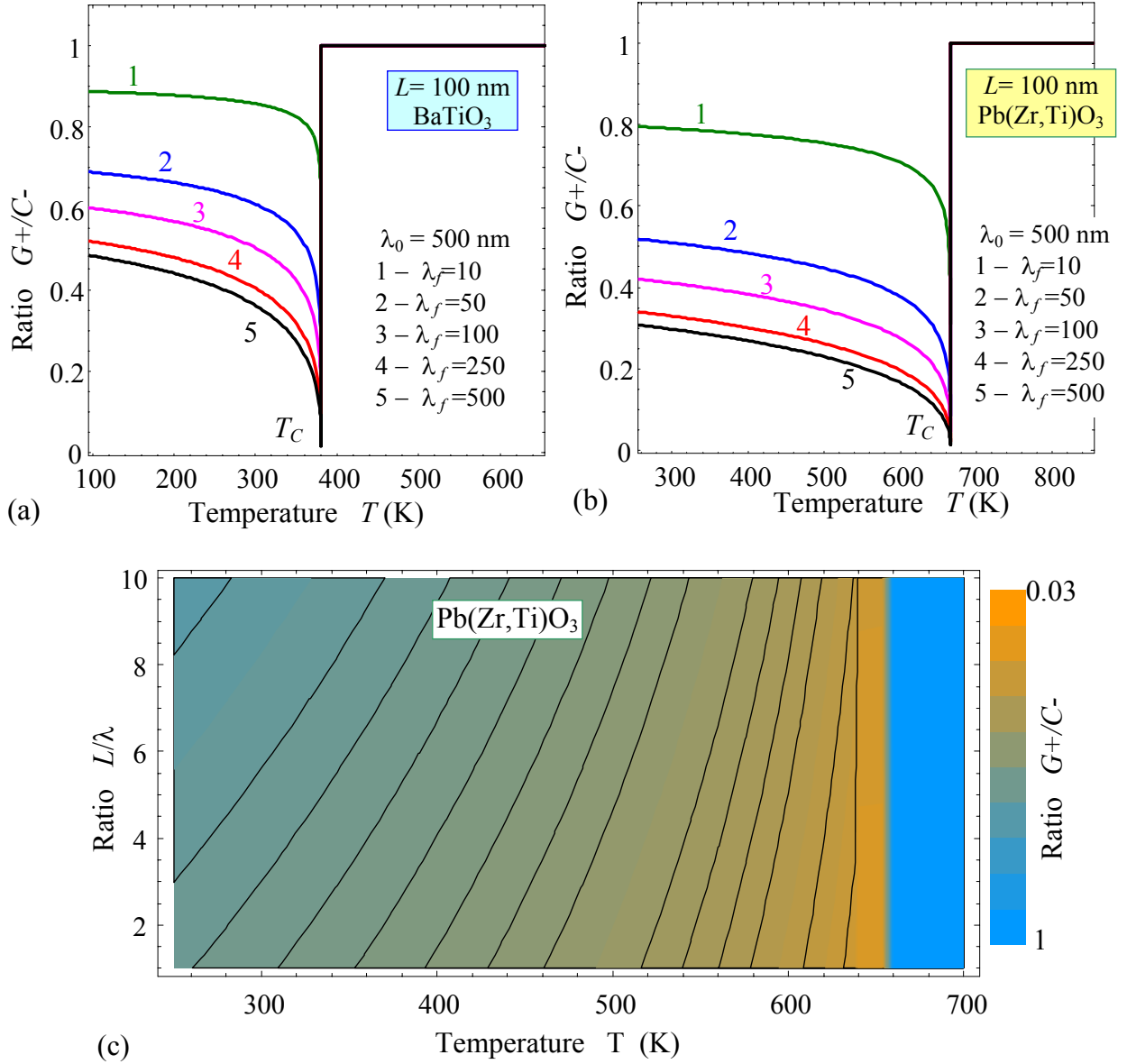

**Figure 4.** The temperature dependence of the ratio $G_+^{total}/G_-^{total}$ calculated for $\lambda_f$= (10, 50, 100, 250, 500) nm (curves 1-5), $\lambda_0 = 500$ nm and $L$=100 nm, BaTiO$_3$ substrate **(a)** and Pb(Zr,Ti)O$_3$ substrate **(b)**. **(c)** Contour map of the ratio $G_+^{total}/G_-^{total}$ in coordinates "ratio $L/\lambda$ and temperature $T$". Parameters $\lambda_0 = 500$ nm and $\lambda_f$ = 50nm, Pb(Zr,Ti)O$_3$ substrate. Ferroelectric materials parameters are listed in **Table I**.

As one can see from **Figure 4**, rectifying properties of the graphene pnJ become higher it the vicinity of Curie temperature. The revealed effect can be especially important for BaTiO$_3$ substrate, where $T_C$ is within the range of the graphene device operation temperature. However, for *T* higher than Curie temperature the rectifying effect vanishes, because there are no longer FDW, which create pnJ in a graphene channel.

### IV. Summary

We have evolved the theory of the conduction of pnJ in graphene channel, caused by FDW, proposed in [14], for the case of arbitrary current regime; from ballistic to diffusive one. We obtained analytical expression for the ratio $G_+^{total}/G_-^{total}$ of the pnJ conductions for opposite polarities of voltage, applied to the source and drain electrodes of the channel, for different ratios of the graphene channel length *L* to electron mean free path λ, and for different values of ferroelectric substrate permittivity. We have demonstrated, that small values of $G_+^{total}/G_-^{total}$ (0.1 and smaller), which correspond to efficient rectifier based on pnJ in graphene, can be obtained for the ferroelectrics with high permittivity and ratios of *L*/λ~1 or smaller. However, for ferroelectrics with extremely high permittivity >>100, the ratio $G_+^{total}/G_-^{total}$ can be essentially smaller then unity for the case of a pronounced diffusive regime of current as well. This makes high-epsilon ferroelectric substrates, such as relaxors and Pb(Zr,Ti)O$_3$, excellent candidates for the creation of new generation excellent rectifiers based on the pnJ in graphene.

Temperature dependence of $G_+^{total}/G_-^{total}$ ratio is determined by two factors: the dependence of ferroelectric permittivity and polarization (which, in its turn, effects the Fermi energy in graphene and the electron mean free path) on the temperature *T* near Curie phase transition. As we have demonstrated within LGD approach, rectifying properties of the graphene pnJ become better in the vicinity below Curie temperature (this effect can be especially important for BaTiO$_3$ substrate, where $T_C$ is within the range of the graphene device operation temperature). However, for *T* higher than Curie temperature the rectifying effect vanishes, because there are no longer FDWs, which create pnJ in graphene channel.

Finally we underline that a proposed theory was build in WKB approximation in assumption of comparatively smooth pnJ, $k_F d > 1$. Because of the large value of ferroelectric permittivity (>5000 for ferroelectric relaxor, 500 for Pb(Zr,Ti)O$_3$, 120 for BaTiO$_3$, 50 for LiTaO$_3$ and 36 for LiNbO$_3$ at room temperature) this criteria is valid for all the range of systems under consideration.

The other essential restriction, used above, was a small voltage limit, which enables an electron to start at source electrode and to reach finally drain electrode without any acts of inelastic scattering. This limit is accurately valid for a wide range of real devices; however, their operation outside this range needs special examination.